# A Matrix Completion Approach to Linear Index Coding Problem

Homa Esfahanizadeh+, Farshad Lahouti+*, Babak Hassibi*

+School of Electrical & Computer Engineering, College of Engineering, University of Tehran
*Department of Electrical Engineering, California Institute of Technology

*Abstract*— In this paper, a general algorithm is proposed for rate analysis and code design of linear index coding problems. Specifically a solution for minimum rank matrix completion problem over finite fields representing the linear index coding problem is devised in order to find the optimum transmission rate given vector length and size of the field. The new approach can be applied to both scalar and vector linear index coding.

## I. INTRODUCTION

An index coding problem [1] arises when a single source intends to communicate to a number of receivers over a rate-limited noiseless broadcast channel. The sender has a number of messages and each receiver desires a specific subset of messages, while having another subset as side information. A single encoding of the messages can be transmitted in each channel use. The objective is to design an encoding scheme with minimum number of channel uses that satisfies all clients.

Research interest on this seemingly simple problem has been recently fueled by results demonstrating its relation to a number of fundamental problems. In particular, it is shown in [5][6] that network coding for multiple unicast is equivalent to a properly constructed index coding problem. The equivalence of topological interference management problem with index coding is investigated in [7][8].

In [2], the index coding problem is analyzed based on graph theoretical approaches and it is shown that scalar linear index coding problem is equivalent to a rank minimization problem over a finite field. It is argued in [3][4] that the optimal length of a linear index coding problem is hard to identify or approximate even for scalar linear index codes.

Graph theoretical approaches to bound the performance of index coding is reported in e.g., [2][8][9]. In this paper we propose an algebraic solution for constructing a vector linear index code and identifying or approximating the optimal rate for an index coding problem by formulating it as a minimum rank matrix completion problem over finite fields.

Despite substantial interest in minimum rank matrix completion over Reals in recent years, the same problem over finite fields has rarely been studied. To the best of our knowledge the only related prior work on this topic is reported in [10]. The algorithm in [10] provides a matrix completion solution of a given rank, when a complete sub-matrix with the same rank is available and the matrix completion solution is unique. The approach relies on identifying the complete sub-matrix by exhaustive search and completing the incomplete rows (columns) in its row (column) subspace in an iterative manner and by search.

In this paper, we propose a new algorithm for matrix completion with minimum rank over finite fields. In the new algorithm, first the complete sub-matrix of highest rank is identified using a heuristic scheme of polynomial complexity, and then row (column) projection is used to expand the complete sub-matrix iteratively. The goal of projection step is to find possible completions of an incomplete row or column such that they are in the span of the complete sub-matrix. During projection, block erasure channel decoding approaches are utilized for enhanced computational efficiency. The row and column projections are administered over a decision tree, which efficiently manages the completion in specific cases when multiple completion solutions are to be examined. This also accommodates the cases where the existing complete sub-matrix is of a smaller rank compared to the (possibly unknown) optimum rank of the prospective complete matrix. As we shall demonstrate, the proposed algorithm provides a solution to identify or approximate the optimum rate for a general linear index coding problem, and to design the associate coding scheme for a given vector (block) length and field size.

The structure of this article is as follows. In Section II, the index coding problem and its associated minimum rank matrix completion problem over finite fields are elaborated. Section III presents the proposed algorithm for matrix completion and section IV presents the simulation results.

## II. PRELIMINARIES

### A. System Model

An instance of the index coding problem is denoted by $\mathcal{I}(X,R)$, where $X = \{x_1,...,x_{|X|}\}$ is the collection of messages at the source ($|.|$ shows the cardinality of a set, $x_i = (x_{i1},...,x_{in}) \in \Sigma^n$, and $\Sigma = \{0,1,...,q-1\}$), and $R$ is the set of receivers. Each receiver $r \in R$ is specified by a pair $(x_r, H_r)$, where $x_r \in X$ is the message which the receiver wishes to decode and $H_r \subset X \setminus x_r$ is the set of messages available at the receiver as side information.

A $C(n,q)$ index code with $l$ transmissions (known as the length of index coding) is a function $f:\left(\Sigma^n\right)^{|X|} \to \Sigma^l$

$$f(X) = \left(f^1(X),\ldots,f^l(X)\right); \quad f^i:\left(\Sigma^n\right)^{|X|} \to \Sigma \quad 1 \le i \le l$$

such that for each receiver $r$, there exists a decoding function $g_r : \Sigma^{l+n|H_r|} \to \Sigma^n$ that $g_r\left(f(X), H_r\right) = x_r$. The transmission rate of an index code $C(n,q)$ is denoted by $\lambda(n,q) = n/l(n,q)$. If $l(n,q)$ is the smallest such integer, $C(n,q)$ is an optimal index code and $\lambda(n,q)$ is the optimal rate for the problem. The objective is to design an encoding scheme with minimum possible length. A scalar or vector (block) index code is identified when $n=1$ or $n>1$, respectively.

In this paper, we are interested in linear index coding in which the encoder, $f(X)$, and decoders, $g_r\left(f(X),H_r\right)$, are linear in their variables. We assume without loss of generality that each receiver needs exactly one message. Obviously any receiver desiring more than one message can be replaced by a number of receivers each requesting one message with the same side information.

*B. Index Coding and Matrix Completion*

Given an instance of linear index coding problem $\mathcal{I}(S,R)$ with block length $n$ and field size $q$, we construct an associated incomplete matrix $M$ in which:
(1) Number of rows, $a$ = number of messages
(2) Number of columns, $b$ = number of receivers
(3) Each element $m_{ij}$ of $M$ may be 1, 0, or $X$ (erasure).

If the receiver $j$ desires the message $i$, $m_{ij} = 1$. If the receiver $j$ has the message $i$ as side information, we consider $m_{ij} = X$ which reflects the fact that the message may be removed from the received signal if it arrives. Finally, if receiver $j$ neither desires message $i$, nor has it as side information, $m_{ij} = 0$, which shows that no interference from this message can be excluded from the desired message. Figs. 1.(a) and 1.(b) depict an index coding problem, and the structure of its equivalent incomplete matrix for binary scalar realization ($n=1$, $q=2$).

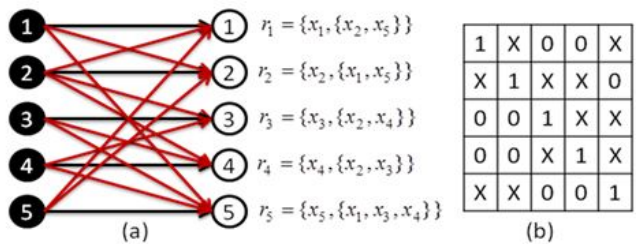

Fig. 1. (a) An index coding problem: the messages are shown by black circles, and the receivers are shown by white circles. A black or red arrow connects a receiver to its desired source or what it has as side information. Fig. 1. (b) corresponding incomplete matrix

Indentifying unknown elements of the incomplete matrix in a way that the completed matrix has the minimum possible rank is equivalent to designing a binary scalar linear index code with minimum number of channel uses. A solution to the minimum rank matrix completion can be translated into a valid encoding scheme for the corresponding index coding problem. Suppose $M^*$ is a solution of the matrix completion problem with rank $k^*$, then there exist $k^*$ independent columns and all other columns are linear combinations of them. Over $i$ th transmission, a combination of the messages specified by the $i$ th independent column is transmitted, each receiver can decode its desired message by linear combining of the received message according to the rule that its corresponding column is formed from the independent columns. As a result, the number of independent columns is the length of the index code, and minimum rank matrix completion is an approach to design a code with minimum possible length over a finite field.

*C. Vector Linear Index Coding*

It is shown that in general, scalar linear index coding is not optimal, and vector linear index coding may be advantageous in certain cases [11]. Given the (incomplete) matrix of a scalar linear index coding problem, we may construct a matrix that corresponds to the associated time extended (vector) linear index coding problem. This is accomplished as follows:
- A "1" is replaced by identity matrix $I_{n*n}$
- A "0" is replaced by zero matrix $0_{n*n}$
- An erasure is replaced by an all erasure matrix of size $(n,n)$

These rules are based on the fact that each message in the $GF(q^n)$ can be substituted with $n$ messages in $GF(q)$. Each receiver desiring a message in $GF(q^n)$ can be substituted with $n$ receivers with same set of side information each wanting a message in $GF(q)$.

### III. ALGORITHM FOR MINIMUM RANK MATRIX COMPLETION

As stated, we aim at completing a matrix over a finite field with minimum rank. The proposed solution consists of two main steps. In the first step, maximal complete sub-matrix of highest rank is identified, and in the second step, row and column projections are used iteratively by means of block erasure decoding techniques until the matrix is completed. Projection means attempting to complete one or more incomplete rows (columns) at each iteration in a way that they are in the span of the current complete sub-matrix. For the example of Fig. 1, Fig. 2 demonstrates the incomplete matrix and the two mentioned steps.

| 1 | X | 0 | 0 | X |
|---|---|---|---|---|
| X | 1 | X | X | 0 |
| 0 | 0 | 1 | 1 | X |
| 0 | 0 | X | 1 | X |
| X | X | 0 | 0 | 1 |

(a)

| 1 | X | 0 | 0 | X |
|---|---|---|---|---|
| X | 1 | X | X | 0 |
| 0 | 0 | 1 | 1 | X |
| 0 | 0 | X | 1 | X |
| X | X | 0 | 0 | 1 |

(b)

| 1 | X | 0 | 0 | X |
|---|---|---|---|---|
| X | 1 | X | X | 0 |
| 0 | 0 | 1 | 1 | X |
| 0 | 0 | X | 1 | X |
| X | X | 0 | 0 | 1 |

(c)

Fig. 2. (a) identification of maximal complete sub-matrix of highest rank, (b) row projection, (c) column projection for matrix completion

In projection step, there might be none or more than one solution for completing rows (columns) such that they are in the row (column) span of the complete sub-matrix. If no solution exists the rank of the matrix is to be increased. In both cases and in order to administer ways of completing the matrix, a decision tree is formed.

In the sequel, we first present the proposed algorithm for identifying the maximal complete sub-matrix of highest rank. Next, the structure of the decision tree is elaborated. Subsequently, we present the proposed row or column projection scheme based on erasure decoding, and then a pruning method is introduced to manage the growth of the tree.

### A. Identification of Maximal Complete Sub-matrix of Highest Rank

Consider the incomplete matrix $M_{a \times b}$, and a complete sub-matrix $M(I_1, I_2)$ identified by rows $I_1 \subset \{1, 2, \ldots, a\}$, and columns $I_2 \subset \{1, 2, \ldots, b\}$ and of rank $k$. Given the matrix $M$, we would like to find the maximal sub-matrix of highest rank. In other words, we wish to find $I_1, I_2$, and $k$ such that:

(1) $M(I_1, I_2)$ is complete.
(2) No other complete sub-matrix could be found with a rank higher than $k$.
(3) $M(I_1, I_2)$ is not sub-matrix of any other matrices having above properties.

The proposed algorithm starts with an initial complete sub-matrix and tries to improve it iteratively. Consider an iteration in which improving the row set, $I_1$, is intended. Let $\bar{I}_1 = \{1, \ldots, a\} \setminus I_1$ be the complement set of $I_1$. The algorithm checks each row in $M(\bar{I}_1, I_2)$ with preference to rows with smaller number of erasures. Assume $i \in \bar{I}_1$ is selected and $I_1' = I_1 \cup \{i\}$; next, all incomplete columns in $M(I_1', I_2)$ are eliminated resulting new column set $I_2'$. If rank of $M(I_1', I_2')$ is higher than $M(I_1, I_2)$, the index sets are updated to new sets $I_1'$ and $I_2'$. A similar approach for column set improvement is taken in turn. These operations are repeated iteratively until no changes are detected over $N$ iterations. The related algorithm is presented in Fig. 3.

---
Algorithm 1
Input: incomplete matrix $M_{a \times b}$
Output: $I_1$, $I_2$ and $k$
Initialize: $I_1 = \{1, \ldots, a\}, I_2 = \Phi, k = 0, N = 100$
Until (No change detected over $N$ iterations in sequence)
    Set $th = |I_1| / (|I_1| + |I_2|)$
    If $rand([0\ 1]) > th$
        Choose row set for improvement
    Else
        Choose column set for improvement
    End
    Set $I_1'$, $I_2'$
    If ($rank(M(I_1', I_2')) > k$)
        Update $I_1$, $I_2$ and $k$
    End
End

---

Fig. 3. Algorithm for maximal complete sub-matrix identification

While it is preferred to identify the maximal complete sub-matrix of highest rank within the matrix $M$, we can proceed to the subsequent steps for matrix completion even with a sub-optimal choice. As such, we may limit the number of iterations in Algorithm 1 to $N$ iterations. Indeed, the choice of $N$ provides a complexity trade-off between the two steps of the proposed matrix completion algorithm. The computational complexity of Algorithm 1 in each iteration is mainly due to computing the rank, hence the algorithm is of polynomial complexity $O(\min(a, b)^3)$.

### B. Decision Tree for Matrix Completion

Each branch in the decision tree is a possible way for completing the matrix with a given rank. The single initial branch in the tree corresponds to the initial incomplete matrix. Components of each branch are:
- Partially completed matrix: $M$
- Index sets showing complete sub-matrix: $I_1, I_2$
- Rank of the complete sub-matrix in the branch: $k$
- Projection direction in the branch

Fig. 4 depicts a possible realization for the decision tree.

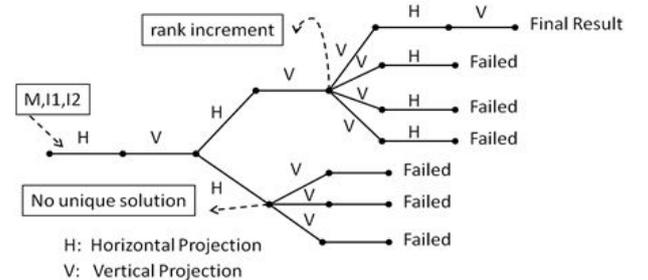

H: Horizontal Projection
V: Vertical Projection
Fig. 4 A sample structure for the decision tree

Consider the horizontal projection in a branch. The vertical projection may be carried out in a similar fashion. There are three possible situations:

1. There are one or more incomplete rows that may be completed uniquely in the subspace spanned by the rows of the current complete sub-matrix. We complete these rows, update the complete sub-matrix and switch the projection direction over the next branch.
2. There is no row that may be completed uniquely as linear combinations of rows of the current complete sub-matrix. This means that all incomplete rows have more than one solution for completion. We choose the incomplete row with minimum possible solutions and analyze their consequent matrix completion over multiple subsequent branches in the alternate direction of projection.
3. There is at least one incomplete row that may not fit within the subspace spanned by the current complete sub-matrix. In this case, the rank of the solution is to be increased. If the increased rank is larger than the rank of the previously completed branches, the current branch is eliminated; otherwise, the possible solutions are examined over multiple subsequent branches.

In each iteration, the branch with maximum opportunity to be the final solution is selected for the projection. This opportunity is quantified by a metric defined by the ratio of the completion percentage of the matrix with respect to the rank of its complete sub-matrix. These projections are done iteratively for branches until they are eliminated or their matrices are fully completed. Then the branch with minimum achieved rank identifies the solution. The related algorithm is presented in Fig. 5.

---
Algorithm 2
Input: incomplete matrix $M_{a \times b}$
Output: Complete matrix with minimum possible rank
Minimum achieved rank= $\min(a,b)$ )
Find maximal complete sub-matrix $M(I1,I2)$ using Algorithm1
While (incomplete branches>0)
    Choose the branch with maximum opportunity
    Perform projection in the proper direction
    Based on projection results add more branches or
        eliminate current one if necessary
    If (matrix is complete in this branch)
        If (achieved rank<minimum achieved rank)
          Update minimum achieved rank
        End
    End
End

---
Fig. 5 Algorithm for projection-based completion algorithm

### C. Projection using Erasure Decoding Techniques

For efficient row (column) projection, we use erasure decoding techniques. The set of all independent rows in the complete sub-matrix is the generator matrix $G$, of a block linear code, and an incomplete row is considered as a codeword generated from this block linear code which is received with some erasures. Having generator matrix $G$, parity matrix $H$, can be constructed ($GH^T = 0$). Then the rows in $H$ specify all the constraints on a vector for being a codeword generated from this block linear code.

Consider an incomplete matrix $M$ with colored complete sub-matrix in Fig. 6.(a). The generator matrix is first constructed by selecting independent rows in the complete sub-matrix and is converted to the row reduced echelon form (Fig. 6.(b)) to efficiently obtain the parity check matrix. The row reduced echelon form of a matrix with rank $k$ has this property that contains an identity matrix $I_{k*k}$ (highlighted in the generator matrix of Fig. 6.(c)). If the generator matrix is in systematic format $G = [I | P]$, then we can construct parity check matrix easily as $H = [P^T | I]$ (Fig. 6.(c)).

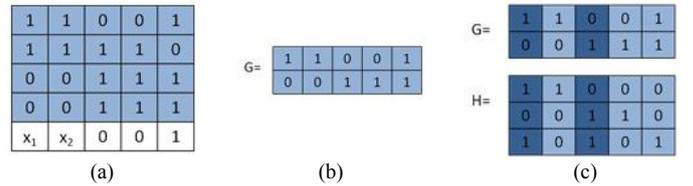

Fig. 6. (a) an incomplete matrix with highlighted maximal complete sub-matrix, (b) row reduced echolon form of complete sub-matrix as a generator matrix, (c) parity check matrix constructed from the generator matrix

The parity check matrix determines all constraints on a codeword $[x_1 x_2 x_3 x_4 x_5]$ generated by this block linear code:
$$x_1 \oplus x_2 = 0, x_3 \oplus x_4 = 0, x_1 \oplus x_3 \oplus x_5 = 0$$
By examining these constraints on the incomplete row, it is found that $x_1 = x_2 = 1$, and the row is completed.

Converting a matrix to its row reduced echelon form has a polynomial complexity $O(|I_1||I_2|^2)$. Constructing parity matrix from the generator matrix has a complexity $O(|I_1||I_2|)$. In a single horizontal projection, the number of incomplete rows is $m - |I_1|$, and the maximal complete sub-matrix with rank $k$ defines $|I_2| - k$ constraints on the incomplete rows. In order to check each constraint on an incomplete row, maximum $|I_2|$ multiplication is needed. The total complexity of horizontal projection is less than $O((m - |I_1|)(|I_2| - k)|I_2| + |I_1||I_2|^2 + |I_1||I_2|)$ which is bounded above by $O(\max(a,b)^3)$.

The growth rate of the tree depends upon the difference between the minimum possible rank $k^*$, and rank of the initial maximal complete sub-matrix $k$. The growth of the tree due to a rank increment is bounded above by $q^\beta$, where $\beta$ is the maximum number of erasures in the rows or columns, and the maximum number of rank increments in a path of the tree is $k^* - k$. Furthermore, the growth due to the multiple allowed combinations for completing a row or column is bounded above by $q^{\max(a,b)-k}$ for each projection because each constraint removes one degree of freedom of erasures (say in a completely erased row) and the total number of erasure is bounded above by $\max(a,b)$. This is still bearable since in most cases, there is an incomplete row (column) with a unique solution (no new branch) or a small number of possible solutions and other rows (columns) with higher number of solutions are not considered in the projection. However, if the size of the tree is noticeably large, the sub-optimum algorithm introduced in the next subsection is useful.

*D. Pruning and Sub-optimum Algorithm*

For large incomplete matrices, which correspond to scenarios with large number of messages, receivers or vector (block) sizes, we propose a pruning scheme to manage the complexity. Specifically, whenever the number of branches goes beyond a preset threshold, the branches with smaller metric (see Section III.B) are pruned.

Even if in some situations, the optimum branch is pruned from the tree; the final low rank solution is still useful since it is a code design with small number of transmissions and can be a good approximate for the optimum transmission rate.

IV. SIMULATION RESULTS

For the index coding problem in Fig.1, the completed matrix based on the proposed algorithm is of rank 2, with all erasures identified as '1'. The corresponding index code is constructed by selecting two independent columns, e.g. $y_1 = x_1 + x_2 + x_5$, $y_2 = x_2 + x_3 + x_4$. Accordingly, all receivers can decode their desired message by using these two transmissions and their side information. The optimal transmission rate is 1/2.

The second example is a 7-multiple unicast index coding. It is proven in [8] that any topological interference management problem can be formulated as a linear index coding problem. An instance of the topological interference management problem and its associated index coding are depicted in Fig. 7. This problem has been solved in [8] by a graph theoretical approach (without constraining the block length and the field size) leading to an optimal transmission rate of 2/5.

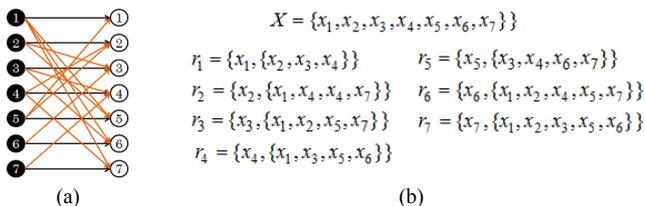

Fig 7. (a) A 7-multiple unicast topological interference management problem, (b) corresponding index coding problem

| $n$ | Size | Erasure | Initial Rank | Pruning Threshold | Number of Tests | Achieved Rank | Avg. Runtime |
|---|---|---|---|---|---|---|---|
| 1 | 7×7 | 59% | 2 | ∞ | 10 | 3 | 0.9 sec |
| 2 | 14×14 | 59% | 4 | ∞ | 10 | 5 | 21 min |
| | | | | 2000 | 10 | 5 | 13 min |
| | | | | 500 | 10 | 7 times 5 | 11 min |
| | | | | | | 3 times 6 | 32 min |

Table 1. Matrix completion for a 7-multiple unicast index coding over binary field with different levels of pruning. The results are obtained on Matlab 2013, CPU Core i5, and 4GB of RAM.

We solved the problem for the case of $q = 2$ and block length $n = 1, 2$ with the proposed algorithm, and reported the results in Table 1. As evident, with $n=1$ a transmission rate of 1/3 and with $n = 2$ a rate of 2/5 is achieved over the binary field. The results indicate that pruning could speed up the computations. Interestingly, achieving the optimum transmission rate (and obtaining the associated index code) is feasible over the binary field with $n = 2$. And there is no need for a larger field size.

Since the index code achieving this optimal transmission rate is not unique, and the initial complete sub-matrix has a lower rank in comparison to the minimum achieved rank, the scheme in [10] fails to complete the matrix.

V. CONCLUSIONS

In this paper, we proposed a constructive approach for code design and rate analysis of linear index coding problems. The index coding problem was formulated as a minimum rank matrix completion problem over finite fields for which an algebraic solution was presented. The proposed approach can approximate or identify the optimum rate of general linear index coding problems with specified accuracy.